\documentclass[10pt, onecolumn, draftcls]{IEEEtran}

\IEEEoverridecommandlockouts

\UseRawInputEncoding

\usepackage[greek,english]{babel}

\usepackage{graphicx,cite,acronym,amsmath,amssymb,amsfonts,color,subfigure,floatflt,stfloats}
\usepackage{epsfig,epstopdf,psfrag}
\usepackage[ruled,vlined]{algorithm2e}
\SetKwInput{KwInput}{Input}                
\SetKwInput{KwOutput}{Output}  

\usepackage[LGR,T1]{fontenc}

\usepackage[table]{xcolor}
\usepackage{bm}
\usepackage[level=3]{wgroup_message}  

\usepackage[cmintegrals]{newtxmath}
\usepackage{graphics,hhline,array,cite,bbm}

\usepackage{latexsym}
\usepackage{theorem}
\usepackage{times}
\usepackage{makeidx}

\DeclareGraphicsExtensions{.png,.jpg,.eps}

\acrodef{IIoT}{industrial Internet-of-things}

\acrodef{SVD}{singular value decomposition}

\acrodef{PSWF}{prolate spheroidal wave function}

\acrodef{CR}{channel response}

\acrodef{BS}{base station}

\acrodef{MS}{mobile station}

\acrodef{UE}{user equipment}

\acrodef{MIMO}{multiple-input multiple-output}

\acrodef{RIS}{reconfigurable intelligent surface}

\acrodef{IRS}{intelligent reconfigurable surface}

\acrodef{LIS}{large intelligent surface}

\acrodef{MIS}{medium intelligent surface}

\acrodef{SIS}{small intelligent surface}

\acrodef{DoF}{degrees-of-freedom}

\acrodef{AF}{amplify \& forward}

\acrodef{DF}{detect \& forward}

\acrodef{JF}{just forward}

\acrodef{CSI}{channel state information}

\acrodef{RV}{random variable}

\acrodef{i.i.d.}{independent, identically distributed}

\acrodef{PSD}{power spectral density}

\acrodef{PDF}{probability distribution function}

\acrodef{CDF}{cumulative distribution function}

\acrodef{ch.f.}{characteristic function}

\acrodef{AWGN}{additive white Gaussian noise}

\acrodef{RSSI}{received signal strength indicator}

\acrodef{SNR}{signal-to-noise ratio}

\acrodef{LRT}{likelihood ratio test}

\acrodef{GLRT}{generalized likelihood ratio test}

\acrodef{GML}{generalized maximum likelihood}

\acrodef{LOS}{line-of-sight}

\acrodef{NLOS}{non-line-of-sight}

\acrodef{GDOP}{geometric dilution of precision}

\acrodef{GPS}{Global Positioning System}

\acrodef{FIM}{Fisher information matrix}

\acrodef{PEB}{position error bound}

\acrodef{WSN}{Wireless Sensor Network}

\acrodef{MAC}{medium access control}

\acrodef{RSS}{received signal strength}

\acrodef{RTT}{round-trip time}

\acrodef{MIMO}{multiple-input multiple-output}

\acrodef{MF}{matched filter}

\acrodef{ED}{energy detector}

\acrodef{ML}{maximum likelihood}

\acrodef{NL}{nonlinear}

\acrodef{MSE}{mean square error}

\acrodef{RMSE}{root mean square error}

\acrodef{ppm}{part-per-million}

\acrodef{PRP}{pulse repetition period}

\acrodef{ACK}{acknowledge}

\acrodef{UWB}{ultrawide bandwidth}

\acrodef{TNR}{threshold-to-noise ratio}

\acrodef{NLOS}{non line-of-sight}

\acrodef{LOS}{line-of-sight}

\acrodef{LS}{least squares}

\acrodef{IR-UWB}{impulse radio UWB}

\acrodef{FCC}{Federal Communications Commission}

\acrodef{TH}{time-hopping}

\acrodef{PPM}{pulse position modulation}

\acrodef{PAM}{pulse amplitude modulation}

\acrodef{MUI}{multi-user interference}

\acrodef{PDP}{power delay profile}

\acrodef{PPP}{Poisson point process}

\acrodef{DS}{delay spread}

\acrodef{CED}{channel excess delay}

\acrodef{BPZF}{band-pass zonal filter}

\acrodef{SIR}{signal-to-interference ratio}

\acrodef{RFID}{radio frequency identification}

\acrodef{WPAN}{wireless personal area networks}

\acrodef{WWLB}{Weiss-Weinstein lower bound}

\acrodef{DP}{direct path}

\acrodef{MF}{matched filter}

\acrodef{MMSE}{minimum-mean-square-error}

\acrodef{SBS}{serial backward search}

\acrodef{NBI}{narrowband interference}

\acrodef{WBI}{wideband interference}

\acrodef{INR}{interference-to-noise ratio}

\acrodef{CIR}{channel impulse response}

\acrodef{ISI}{inter-symbol interference}

\acrodef{CPR}{channel pulse response}

\acrodef{LRT}{likelihood ratio test}

\acrodef{RADAR}{RADAR}

\acrodef{MUR}{Multistatic RADAR}

\acrodef{MUI}{multi-user interference}

\acrodef{e.m.}{electromagnetic}

\acrodef{EM}{electromagnetic}

\acrodef{CW}{continuous wave}

\acrodef{RF}{radiofrequency}

\acrodef{FCC}{Federal Communications Commission}

\acrodef{EIRP}{effective radiated isotropic power}

\acrodef{RCS}{radar cross section}

\acrodef{BAV}{balanced antipodal Vivaldi}

\acrodef{PRake}{partial Rake}

\acrodef{RTLS}{real time locating system}

\acrodef{CRB}{Cram\'{e}r-Rao bound}

\acrodef{ZZB}{Ziv-Zakai bound}

\acrodef{TOA}{time-of-arrival}

\acrodef{TOF}{time-of-flight}

\acrodef{WSN}{wireless sensor network}

\acrodef{MAC}{medium access control}

\acrodef{RSS}{received signal strength}

\acrodef{TDOA}{time difference-of-arrival}

\acrodef{RF}{radiofrequency}

\acrodef{RTT}{round-trip time}

\acrodef{AOA}{angle-of-arrival}

\acrodef{MF}{matched filter}

\acrodef{ED}{energy detector}

\acrodef{ML}{maximum likelihood}

\acrodef{MUR}{Multistatic radar}

\acrodef{HDSA}{high-definition situation-aware}

\acrodef{RRC}{root raised cosine}

\acrodef{OFDM}{orthogonal frequency division multiplexing}

\acrodef{IF}{intermediate frequency}

\acrodef{PHY}{physical layer}

\acrodef{S-V}{Saleh-Valenzuela}

\acrodef{UHF}{ultra-high frequency}

\acrodef{PR}{pseudo-random}

\acrodef{SoC}{System on Chip}

\acrodef{SoP}{System on Package}

\acrodef{SPMF}{Single-Path Matched Filter}

\acrodef{IMF}{Ideal Matched Filter}

\acrodef{SCR}{signal-to-clutter ratio}

\acrodef{BEP}{bit error probability}

\acrodef{BER}{bit error rate}

\acrodef{WSR}{wireless sensor radar}

\acrodef{HPBW}{half power beam width}

\acrodef{LEO}{localization error outage}

\acrodef{WSS}{wide-sense stationary}

\acrodef{TR}{time-reversal}

\acrodef{WSSUS}{WSS with uncorrelated scattering}

\acrodef{GP}{Gaussian process}

\acrodef{IMU}{inertial measurement unit}

\newcommand{\bolds} {{\bf{s}}}

\newcommand{\boldr} {{\bf{r}}}

\newcommand{\Sr} {\mathcal{S}_{\text{R}}}
\newcommand{\St} {\mathcal{S}_{\text{T}}}

\newcommand{\bphi} {\it \phi}
\newcommand{\bpsi} {\it  \psi}

\begin{document}
\title{Holographic Communication\\ using Intelligent Surfaces}

\author{
\IEEEauthorblockN{Davide~Dardari,~\IEEEmembership{Senior~Member,~IEEE}, Nicol\'o Decarli,~\IEEEmembership{Member,~IEEE}}
\IEEEcompsocitemizethanks{\IEEEcompsocthanksitem 
 D.~Dardari is with the 
   Dipartimento di Ingegneria dell'Energia Elettrica e dell'Informazione ``Guglielmo Marconi"  (DEI), CNIT, 
   University of Bologna, Cesena Campus, 
   Cesena (FC), Italy, (e-mail: davide.dardari@unibo.it). 
   \IEEEcompsocthanksitem 
   N. Decarli is with the IEIIT-CNR at the University of Bologna, Bologna (BO), Italy (e-mail:  {nicolo.decarli}@ieiit.cnr.it).
    }
}

\maketitle
\begin{abstract}
Holographic communication is intended as an holistic way to manipulate with unprecedented flexibility the electromagnetic field generated or sensed by an antenna. This is of particular interest when using large antennas at high frequency (e.g., the millimeter wave or terahertz), whose operating condition may easily fall in the Fresnel propagation region (radiating near-field), where the classical plane wave propagation assumption is no longer valid.
This paper analyzes  the optimal communication involving large intelligent surfaces, realized for example with metamaterials as possible enabling technology for holographic communication. It is shown that  traditional propagation models must be revised and that, when exploiting spherical wave propagation in the Fresnel region with large surfaces, new opportunities are opened, for example, in terms of the number of orthogonal communication channels. 

\end{abstract}



\section{Holographic Communication}

\IEEEPARstart{T}{he} increasing demand of ubiquitous, reliable, fast and scalable wireless services is pushing today's radio technology towards its ultimate limits. The current deployment of the fifth-generation (5G) wireless networks is expected to exploit increasingly \ac{MIMO} techniques and cell densification, in order to serve a large number of users per area with the required throughput. However, for the sixth-generation (6G) wireless networks, even more stringent requirements are set in terms of data-rate, number of users, reliability, with the goal of enabling massively novel applications, for instance, in the fields of \ac{IIoT} and autonomous driving. In this context, a significant increase of the number of antennas is required (massive-\ac{MIMO}), in conjunction with the exploitation of higher frequencies, where larger bandwidth is available \cite{San:19}. 

The use of millimeter wave and terahertz technologies translates into a larger path-loss, which can be partially compensated by antennas densification and large antenna arrays. \cite{RapXinKanJuMadManAlkTri:19}. 
Indeed, the shift towards large antennas and high frequency poses new challenges since traditional models based on the assumption of far-field \ac{EM} propagation  fail, but opens new opportunities at the same time. 
\begin{figure}[t]
\centerline{\includegraphics[width=0.7\columnwidth]{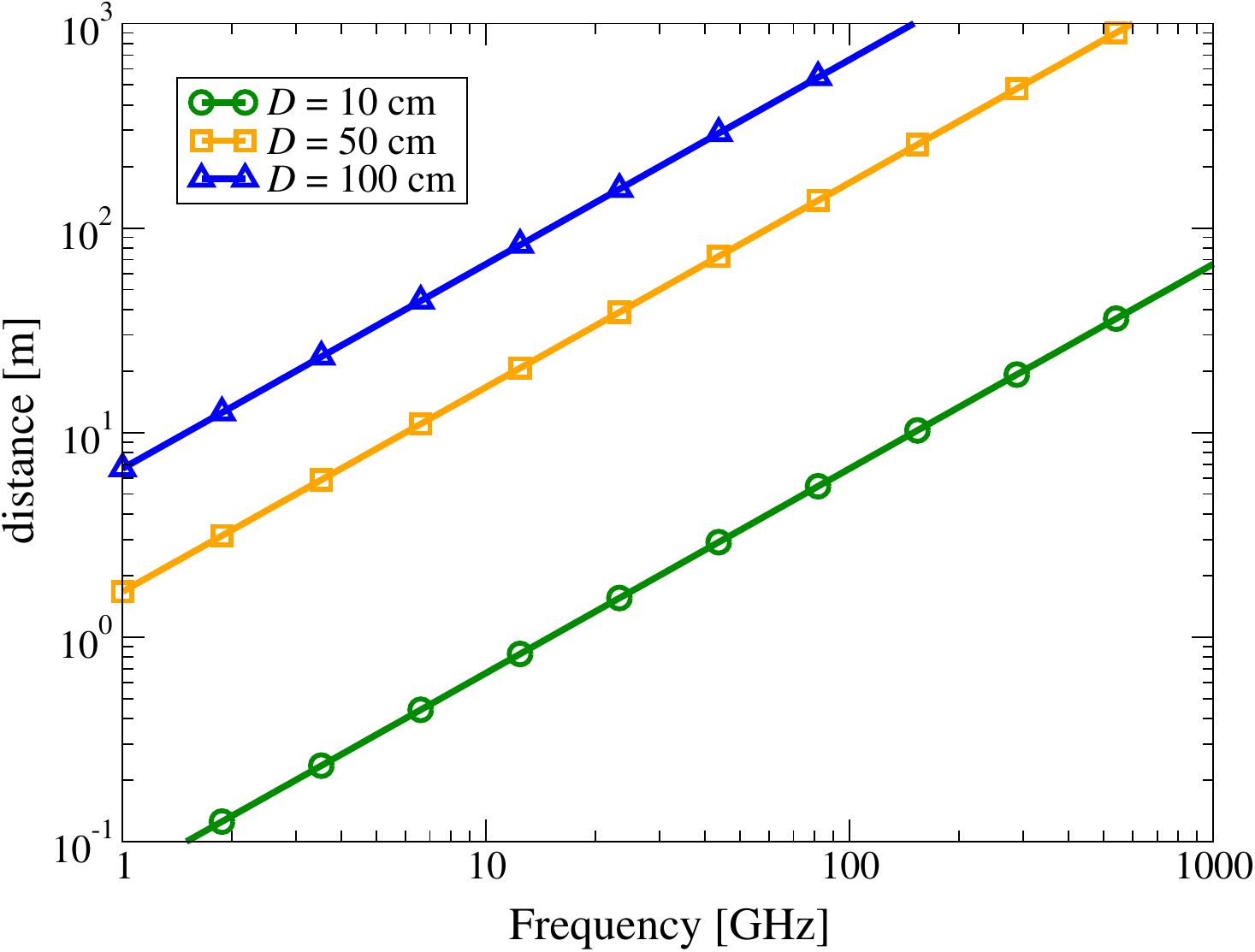}}
\caption{Fraunhofer region boundary as a function of the frequency, for different antenna size $D$. Practical operating distances in the millimeter wave band follow below the Fraunhofer boundary (i.e., in the Fresnel region of the antenna).}
\label{Fig:Fraunhofer}
\end{figure}
In fact, in classical operating conditions, i.e., in  the Fraunhofer region of the antenna, the radio link is much longer than the antenna dimension, so that plane wave propagation is assumed. 
Conversely, when the antenna size becomes comparable to the link distance, operating conditions fall within the Fresnel region in which (radiating) near-field propagation takes place. Fig.~\ref{Fig:Fraunhofer} shows the commonly-assumed boundary between the Fresnel and Fraunhofer regions as a function of the operating frequency, from \ac{UHF} to terahertz, considering different sizes of the antenna. Notably, the lower part of the graph corresponds to the operation in the Fresnel region, whereas the upper part corresponds to the operation in the Fraunhofer region. As it is evident from the figure, when antennas between $10\,$cm and $1\,$m are considered for applications in the millimeter wave band, typical operating distances between 1 and 100 meters are included almost entirely in the Fresnel region, where plane wave approximation of the wave front does not hold anymore, and spherical wave front propagation must be considered instead. 
If, from one side, operation below the Fraunhofer region boundary requires the consideration of new models capable of accounting for this regime, from the other side it opens new unexplored possibilities to enhance the communication performance. 

In this context, to fully exploit the characteristics offered by different \ac{EM} propagation regimes and thus approach the ultimate limits of the wireless channel, the complete control of the \ac{EM}  field generated/sensed by antennas should be reached.  
This is the concept of \textit{holographic communication}. 
The term \textit{holography} derives from the ancient greek  \begin{otherlanguage}{greek}
{olos}\end{otherlanguage}, \emph{holos}, (all), and \begin{otherlanguage}{greek}
{graf\`e}\end{otherlanguage}, \emph{graf\`e}, (writing, drawing) and it literally means ``describe everything''.  The holographic capability of a transmitting antenna in the near-field consists in the possibility to generate any current density distribution on its surface, in order to obtain the maximum flexibility in the design of the radiated \ac{EM} field (amplitude, wave front, polarization...). Similarly, the holographic capability at the receiving antenna side consists in the possibility to weight the impinging electric field according to a desired function, thus manipulating the way the antenna receives the information without the need of any physical modification of its shape.

\begin{figure*}[t]
\centerline{\includegraphics[width=\textwidth]{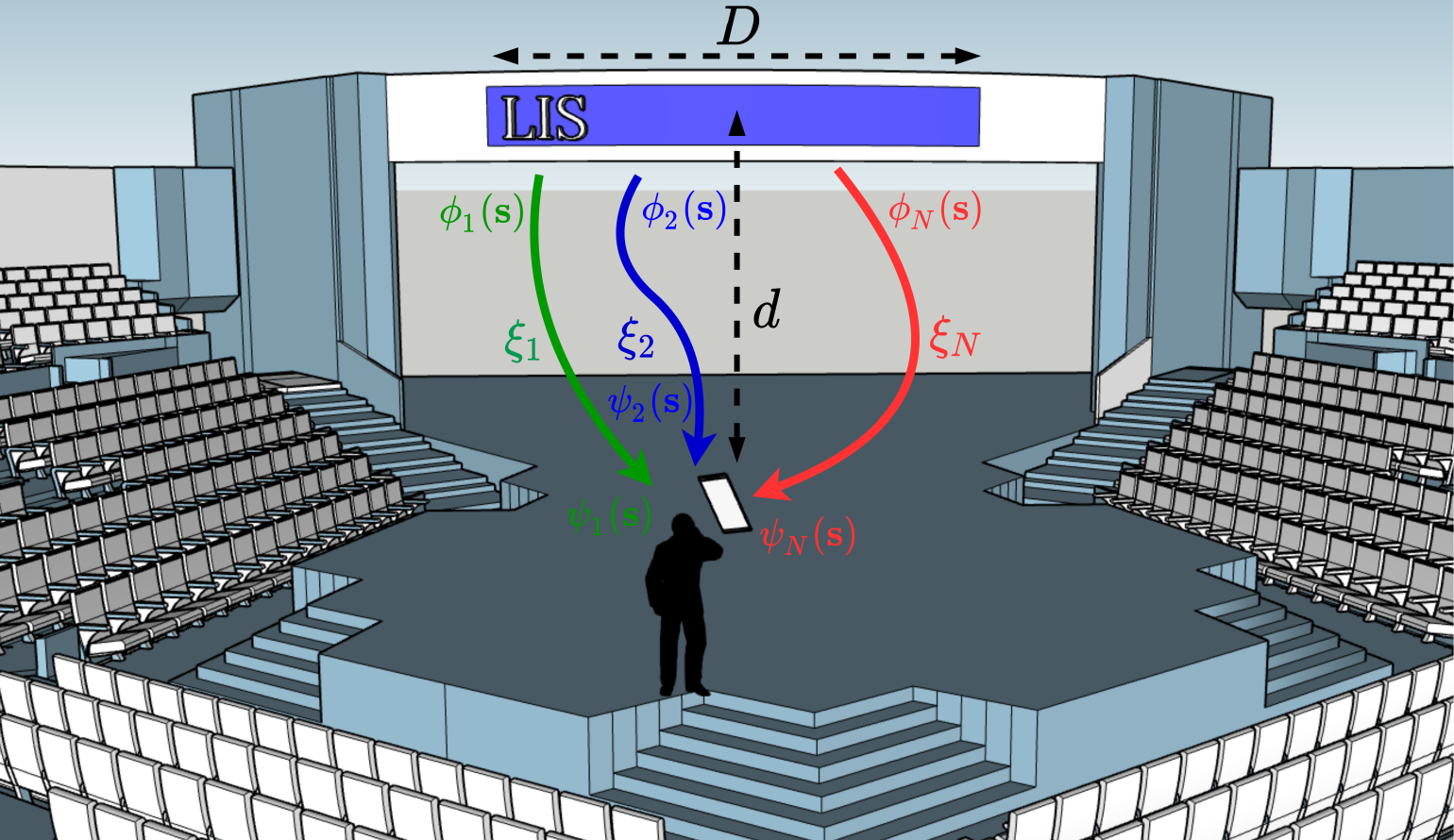}}
\caption{Scenario with a LIS-based antenna communicating with a mobile terminal using multiple \textit{communication modes}.}
\label{Fig:Model}
\end{figure*}

From the technological point of view, metamaterials represent appealing candidates toward the 
creation of \emph{intelligent surfaces}, which can lead to a viable way of realizing highly-flexible antennas \cite{Tre:15, Sil:14,GonMinChaMac:2017,HolKueGorOHaBooSmi:12,Fink:14}. 
In fact, metamaterials enable the manipulation of the \ac{EM} field  or the local control of amplitude and phase reflecting behavior at an unprecedented level, thus enabling the design of specific characteristics in terms of reflection, refraction, absorption, polarization, focusing and steering, when used as reflecting surfaces. 
In the last years, the idea of deploying semi-passive reconfigurable reflecting elements in the environment has attracted a considerable research attention. Such solutions, based on additional entities referred to as \acp{IRS} or \acp{RIS}, are able to create artificial multipath or additional communication channels between a transmitter and a receiver, thus increasing the coverage and the \ac{DoF} of wireless communication \cite{DiRenzoJSAC:20,HuaEtAl:20}. 
As active antennas, intelligent surfaces can be exploited to increase the number of design variables  
allowing to operate directly at \ac{EM} level by processing electromagnetic waves with unprecedented level of flexibility and resolution. When such  antennas are electrically large, they are referred to as \acp{LIS} \cite{HuRusEdf:18}; this definition will be adopted in the rest of the paper. Due to the large size, the radio propagation at millimeter waves and even in the terahertz band may occur in the near-field region of the antenna even at practical distances, and thus traditional assumptions resorting to planar wave front cannot be anymore considered valid. 
The adoption of \acp{LIS} provides high flexibility in network design as well as the potential to achieve the goals of next generation wireless networks, but it also opens several fundamental questions that are still unsolved, such as understanding the theoretical limits of these communication technologies and how to achieve them in practice.

This paper presents the fundamentals of communication with \acp{LIS}. In particular, 
the limits of traditional communication models and the new opportunities offered by the spherical wave front propagation in the Fresnel region are discussed, showing that it is possible to increase significantly the number of orthogonal communication channels between a couple of \acp{LIS}, even in \ac{LOS} and without the need of rich multipath propagation. Finally, the main open research directions in this field are highlighted.

\section{Information-theoretical Optimal Communication between \acp{LIS}}
%
A \ac{LIS} denotes an intelligent metasurface whose size $D$ is  much larger than the operating wavelength, and often comparable to the radio link distance $d$ (see Fig.~\ref{Fig:Model}). When such conditions hold, it is fundamental to account for proper modeling of the \ac{EM} propagation.
 In fact, with \acp{LIS}, classical models for antenna arrays may fail to describe correctly the actual wireless link characteristics in terms of path-loss and number of \emph{communication modes}, i.e., orthogonal \ac{EM} channels, that can be exploited for communication, as they typically assume far-field conditions.
In addition, classical models are based on specific current distributions related to the antenna shape (e.g., dipole, patch, spiral...), without accounting for the flexibility in generating these distributions offered by metasurfaces (holographic capability). Therefore, the models adopted for the description of communication with \acp{LIS} must account for this design flexibility thus relying directly on considerations concerning the transportation of information with \ac{EM} waves in the continuous wireless channel (\ac{EM} information theory).

In order to abstract from the specific implementation of the metasurface, we model the intelligent surface as a continuous array of an infinite number of infinitesimal antennas. The wireless communication exploiting an uncountable infinite number of antennas in a finite space has been recently defined as \emph{holographic \ac{MIMO}} \cite{PizMarSan:19b,HuaEtAl:20}. 
In this manner, implementation-related aspects concerning the mutual coupling among the metasurface elements are not involved in the discussion. 

Optimal communication between \acp{LIS}, considering a continuum of infinitesimal antennas and the continuous wireless channel, can be modeled as the problem of communicating between a couple a spatial regions (or volumes in the case the antenna thickness is not considered negligible). This enables moving away from the classical \ac{MIMO} model of point-defined antennas, which can be considered as a particular case of this general formulation, where the continuous space \ac{EM} channel and  continuous signals (propagating waves) are sampled according to a specific placement of the array elements. Then, communication is viewed as a functional analysis problem depending only on geometric relationships, whose goal is to determine the optimal set of \ac{EM} functions at transmitter and receiver sides to transfer information between the spatial regions. In this manner, the ultimate limits for communication, namely the intrinsic capacity of the continuous-space wireless channel, can be investigated independently of the specific technology and number of antenna elements.

\begin{figure*}[t]
\centerline{\includegraphics[width=\textwidth]{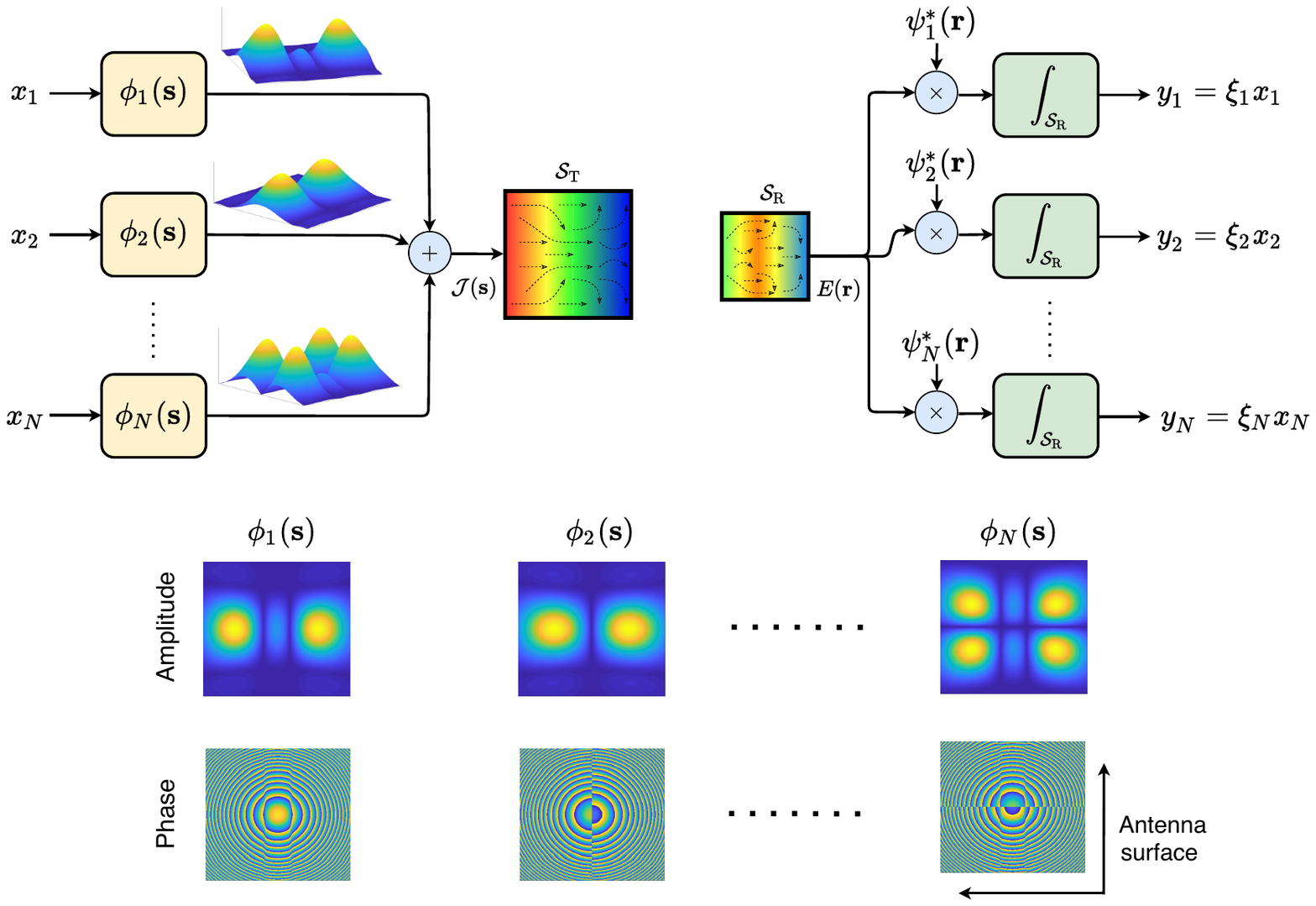}}
\caption{Optimal communication between intelligent surfaces. Example of orthogonal basis function used at the transmitting surface, whose combination gives the current density $\mathcal{J}(\mathbf{s})$. By proper design, orthogonal functions are obtained also on the receiving antenna surface, whose linear combination gives the electric field ${E}(\mathbf{r})$. Parallel orthogonal channels with coupling intensities $\{\xi_n\}$, namely \textit{communication modes}, can be  realized.
}
\label{Fig:ParallelComm}
\end{figure*}

In the pictorial example of Fig.~\ref{Fig:Model}, a \ac{LIS} placed on a large wall (e.g., base station) communicates with an antenna embedded in a portable device in \ac{LOS} channel condition. 
Specifically, with reference to Figs.~\ref{Fig:Model} and \ref{Fig:ParallelComm}, we consider a set of functions $\{\bphi_n(\bolds)\}$ with $\bolds \in \St$, representing any complete basis set of the transmitting  surface $\St$, and a set of functions $\{\bpsi_n(\boldr)\}$ with $\boldr \in \Sr$, representing any complete basis set of the receiving surface $\Sr$, for $n=0,\ldots,\infty$. 
As a consequence, any current density $\mathcal{J}(\bolds)$ on the surface $\St$ can be represented as a proper linear combination of the transmitting basis functions $\{\bphi_n(\bolds)\}$, while a proper linear combination of the receiving basis functions $\{\bpsi_n(\boldr)\}$ describes the electric field $E(\boldr)$ on the surface $\Sr$.
Among the possible choices of the basis sets, an interesting case is that for which the communication operator, which puts into relation the transmit basis function $\bphi_n(\bolds)$ with the receiving basis function $\bpsi_n(\boldr)$, is diagonal, meaning that there is a one-to-one correspondence between  $\bphi_n(\bolds)$ and $\bpsi_n(\boldr)$ through a multiplicative coupling coefficient $\xi_n$, with as large coupling intensity as possible. This is the concept of \textit{communication mode} \cite{Miller:19}.

In order to determine the communication modes, an eigenfunctions problem starting from an electromagnetic description of the continuous wireless channel must be solved; the kernel of the problem is connected to the Green function putting in relationship the effect on a given point of the receiving surface (wave) with an infinitesimal point-wise excitation given on the transmitting surface. The solution to the eigenfunctions problem gives the information-theoretical optimal communication strategy, and the current density excitation $\bphi_n(\bolds)$ at transmitter side will produce an effect (electric field) $\bpsi_n(\boldr)$ at receiver side, without the excitation of the other modes \cite{Dar:J20}. Therefore, multiple parallel and orthogonal channels can be established, as depicted in Fig.~\ref{Fig:ParallelComm}. Notice that, in analogy with \ac{MIMO}, we can see the transmitting function as a form a pre-coding vector, and the receiving function as a form of combining vector, but at \ac{EM} level. 

In practice, since the spatial regions of the antennas are confined, the number of significant (large) eigenvalues is limited. Therefore, 
 the number of communication modes is defined conventionally as the minimum number $N$ of eigenvalues  sufficient to describe the signals within a given level of accuracy, e.g., compared to the noise intensity. 
 This translates into the possibility to represent any current and field distribution as the combination of a limited number of communication modes allowing significant information transfer. 
More specifically, we obtain the input-output representation in terms of $N$ parallel channels (the communication modes) $y_n=\xi_n \, x_n + w_n$, for $n=1,2, \ldots , N$, being $w_n$ the \ac{AWGN}, where the $N$ input data streams $\{ x_n\}$  are associated to the basis functions $\{ \phi_n(\bolds)\}$ in $\St$, and they are recovered at the receiver after the (spatial) correlation of the received electric field with the corresponding basis functions $\{ \bpsi_n(\boldr) \}$ in $\Sr$, as shown in Fig.~\ref{Fig:ParallelComm}.
The eigenfunction decomposition ensures that the current distribution $ \bphi_1(\bolds)$ in $\St$ leads to the electric field $\xi_1 \, \bpsi_1(\boldr)$ within $\Sr$ with the largest intensity $|\xi_1|^2$. The current distribution $\bphi_2(\bolds)$ in $\St$ leads to the electric field $\xi_2 \, \bpsi_2(\boldr)$ within $\Sr$, orthogonal  to $\xi_1 \, \bpsi_1(\boldr)$, with the second largest intensity $|\xi_2|^2$, and so on.
Each pair of functions $\left ( \bphi_n(\bolds),\bpsi_n(\boldr) \right )$  determines a spatial dimension of the system across which one can establish an orthogonal communication which can be exploited to maximize the capacity using the waterfilling approach. 
%
%
%
%
 A large level of coupling means that the generated wave is confined approximately within the space between $\St$ and $\Sr$, thus impressing a current on the receiving surface $\Sr$. Instead, a low level of coupling denotes that the generated wave is mainly dispersed away from the receiver's surface $\Sr$.

Finding the solution of the eigenfunctions problem, thus defining the optimal set of basis functions and coupling coefficients at transmitter and received side, requires extensive and sometimes prohibitive simulations if large surfaces are considered. In particular, a discretization into a fine mesh of the transmitting and receiving regions can be realized, then solving numerically the eigenfunctions problem and applying \ac{SVD} for resorting to the simplest description of the communication operator.
Some analytical results related to specific geometric configurations can be found in \cite{Dar:J20} from which some interesting insights can be obtained. Examples of basis functions for square surfaces are reported in Fig.~\ref{Fig:ParallelComm}.

\section{Communication Modes}

\begin{figure}[t]
\centerline{\includegraphics[width=0.7\columnwidth]{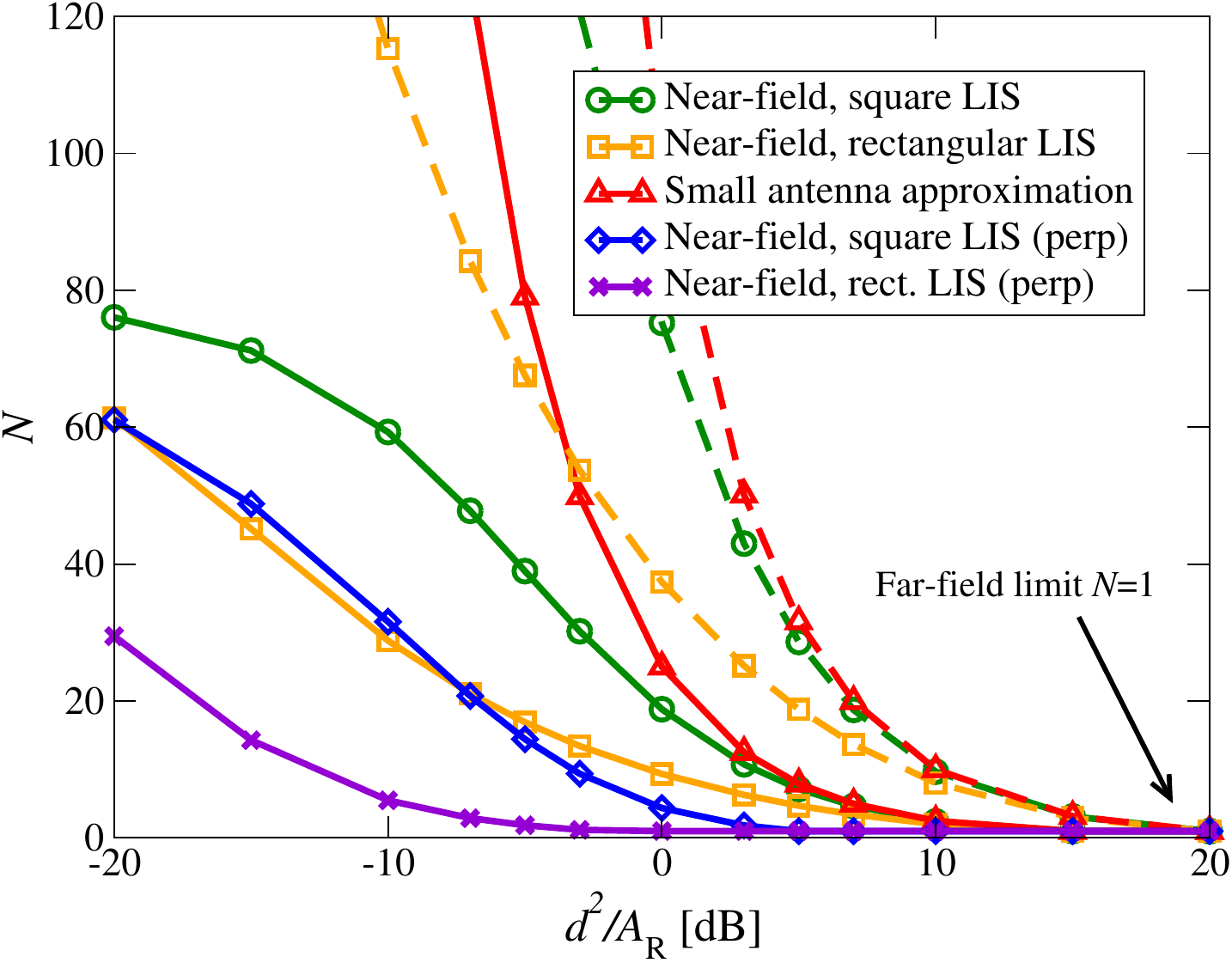}}
\caption{Number of communication modes $N$ as a function of the ratio  $d^2/A_{\text{R}}$, for square and rectangular parallel \acp{LIS}. Continuous lines refer to $28\,$GHz. Dashed lines refer to $60\,$GHz. For $28\,$GHz, parallel and perpendicular orientations among the surfaces are showed. Comparison with the small-antenna approximation (red lines with $\bigtriangleup)$. In the far-field, $N=1$ for LOS conditions.}
\label{Fig:DoF}
\end{figure}

When the size of both surfaces is small compared to the distance $d$ separating them, the number of communication modes $N$ is known from results originally obtained for optical communication by resorting to diffraction theory. Specifically, it is $N=A_{\text{T}}\, A_{\text{R}}/(\lambda\, d)^2$, where  $A_{\text{T}}=|\St|$ and $A_{\text{R}}=|\Sr|$ are the areas of the two surfaces and $\lambda$ is the wavelength. In the far-field, for large $d$, the number of communication modes $N$ is unitary in \ac{LOS} propagation, and the coupling intensity (i.e., the channel gain) scales with $d^2$. This result comes from the analytical solution of the eigenfunctions problem leading to \acp{PSWF} as eigenfunctions, whose eigenvalues are in fact almost equally-sized up to a certain limit, then dropping rapidly to zero \cite{Miller:00}. Specifically, it is based on the paraxial approximation for parallel and collinear surfaces, and it can be exploited also in the Fresnel region of the surfaces, but under the condition that the link distance $d$ remains much larger than the surface's size $D$.
When using \acp{LIS}, the previous result might be no longer valid as it is likely the surfaces work in their Fresnel region and with link distance $d$ comparable with the surface's size $D$, as depicted in Fig.~\ref{Fig:Model}. 

We report now some recent results obtained for communication between a relatively small surface and a \ac{LIS} \cite{Dar:J20}.
Fig.~\ref{Fig:DoF} shows the number of communication modes $N$ that can be obtained in this configuration. In particular, a transmitting surface of area  $A_{\text{T}}=5\times5\,\text{cm}^2$, and frequencies of $28\,$GHz and $60\,$GHz  are considered. Both the cases of parallel and perpendicular orientations among the \acp{LIS} are reported, as well the case of \acp{LIS} with different aspect ratios, i.e., a square \ac{LIS} of area $A_{\text{R}}$, and a rectangular \ac{LIS} of the same area, with 4:1 aspect ratio between height and width. Obviously, due to the reciprocity of the radio medium, the transmitting and receiving surfaces role can be exchanged. The number of communication modes is plotted as a function of the ratio $d^2/A_{\text{R}}$. 
It is evident how a large number of communication modes can be obtained even in the \ac{LOS} scenario (no exploitation of multipath), thus boosting significantly the channel capacity at \ac{EM} level. The result known for the small antenna approximation is also reported for comparison; it can be noticed   that without accounting for the near-field behavior, a large overestimation of the number of communication modes is obtained, especially when a rectangular \ac{LIS} is considered. 

When the distance among the surfaces increases, the number of communication modes decreases and the small surface approximation becomes valid. Under this condition, $N$ is lowered by a factor $\sqrt{d/A_{\text{R}}}$ in the perpendicular  configuration. Differently from the far-field condition, where no communication can be established among perpendicular surfaces, this is not true in the near-field, for which a significant number of communication modes can still be  obtained, especially when the surface's size is comparable with the link distance. 

For very large distances, the limit $N=1$ arises.
On the other extreme cases, i.e., when the \ac{LIS} becomes very large, the number of communication modes for parallel surfaces is given by $\pi A_{\text{T}}/\lambda^2$ \cite{Dar:J20} and it depends only on the area of the transmitting surface, i.e., the smallest surface, normalized with respect to the square wavelength. It is interesting that the same limit arises also for perpendicular surfaces.

\section{Power Scaling Law}

As for the number of communication modes, the near-field characteristics must be accounted also in deriving the power scaling law between a transmitter and a receiver, namely the channel power gain.
When multiple communication modes are excited, the gain is considered as the ratio between the overall received power and the overall transmitted power which equals the sum of the coupling intensity of all the communication modes and it depends only on the geometric configuration \cite{Miller:00}.

\begin{figure}[t]
\centerline{\includegraphics[width=0.7\columnwidth]{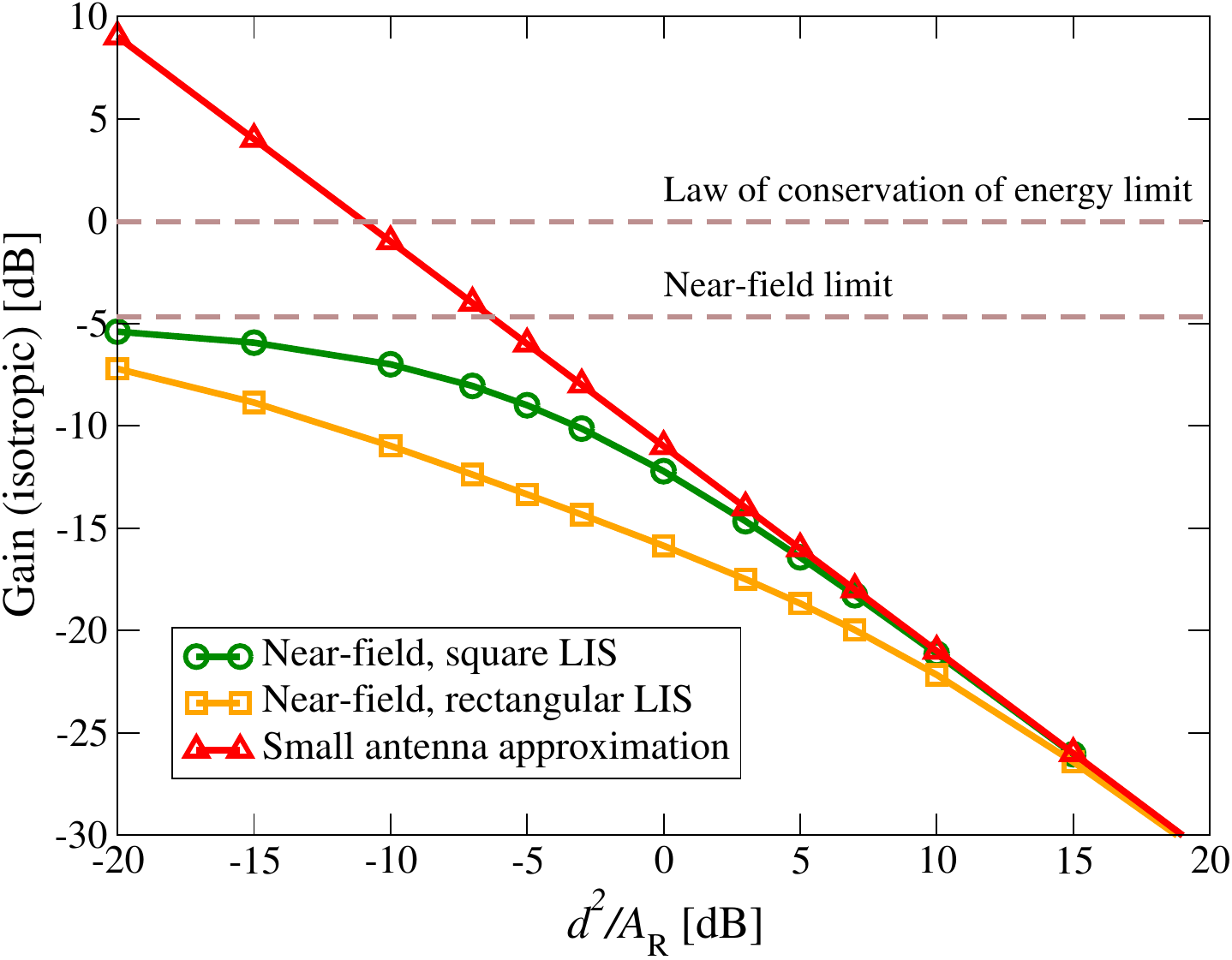}}
\caption{Channel power gain with respect to the ratio $d^2/A_{\text{R}}$, considering a receiving antenna of area $A_{\text{R}}$ and an isotropic antenna at transmitter side. Red lines with $\bigtriangleup$ correspond to the traditional Friis formula, not valid when the link distance $d$ becomes comparable with the surface's size $D$ (violation of the law of conservation of energy).}
\label{Fig:Gain}
\end{figure}

Fig.~\ref{Fig:Gain} reports the channel power gain in the  communication between an isotropic antenna and a square \ac{LIS} in the same configuration as in the previous section. Notice that, due to the assumption of isotropic transmitter, the channel power gain is independent of the operating frequency. 
For comparison, the classical Friis formula, which can be obtained also by considering the small antenna approximation and the solution resorting to the \acp{PSWF} (thus valid in the Fraunhofer region and, in part, in the Fresnel region), is reported.
It is evident that the classical Friis formula is no longer valid when using \acp{LIS} at small distances between the surfaces. This is due to the operating region in the early Fresnel region, with surface's size comparable with the link distance, whose peculiarity is not captured by classical path-loss modelling. In fact, according to the Friis formula, increasing the size of both the transmit and receive surfaces could lead to the increasing of the power gain at any level. This is in contrast with the law of conservation of energy, for which the received power cannot overcome the transmitted one. When one of the two surfaces becomes large, thus working in the near-field, an intrinsic limitation arises, and the maximum channel power gain (with respect to the isotropic antenna) saturates to $1/3$ ($-4.77\,$dB) \cite{Dar:J20}. Thus, with an infinite-size square \ac{LIS}, no more than $1/3$ of the transmitted power can be collected. The limit arises naturally, and it is due to three phenomena happening with \ac{LIS}: when the surface becomes large (i) the distance from the point source to the specific location of the receiving \ac{LIS} can be substantially larger than the link distance (i.e., distance from the point source to the \ac{LIS} center), thus producing higher attenuation; (ii) points far from the \ac{LIS} center exhibits a smaller local effective area as seen by the transmitter; (iii) the polarization mismatch increases for points far from the \ac{LIS} center \cite{BjoSan:20}.  

It is interesting to underline that the result of Fig.~\ref{Fig:Gain} is general, and it does not depend on the operating frequency, but it is  determined only by geometric factors normalized to the wavelength.
The point at which the channel power gain diverges from that predicted by the Friis formula corresponds to a link distance $d$ close to the surface's size $D$, for square surfaces, while it is larger for rectangular surfaces. This is different from the classical definition of the boundary between Fraunhofer and Fresnel regions, which is frequency-dependent. 
Power scaling laws for more complex scenarios involving the presence of \ac{RIS} and relays can be found in \cite{BjoSan:20}.

\section{Research Directions and Conclusion}

Reaching the ultimate limits in wireless communication cannot disregard the physical limits of \ac{EM} propagation, especially when operating at high frequencies and with large antennas.
%
Along this direction, in this paper we have discussed the theoretical implications when using \ac{LIS} as active antennas, by showing the limitations of current models and the opportunities offered when operating in the near-field regime. In fact, while in the far-field only one communication mode can be established in \ac{LOS}, an increased number of communication modes can be obtained in the Fresnel region, and hence  the opportunity to boost the capacity. 
However, the approaching of this potential improvement with practical systems requires solving several theoretical and technological challenges, some of them summarized in the following.

\begin{itemize}

    \item \emph{Metasurface technology:} Despite the considerable advances in metamaterials technology, which have seen the introduction of different solutions for the implementation of \ac{LIS}, such as dynamic metasurface antennas (DMA) based on waveguide-fed metasurface  
    \cite{Eldar:19,SmiOkaPul:17} and multi-beam antennas \cite{GonMinChaMac:2017}, a considerable gap needs to be filled toward fully flexible \acp{LIS}.      
    
    \item \emph{EM-based signal processing:} 
    In perspective, the flexibility offered by metamaterial paves the way  to shift some functionalities that are typically performed in the digital domain directly at \ac{EM} level with the purpose to tackle complexity issues and  reduce significantly the latency, as the processing would be realized at the speed of light \cite{Sil:14}. 
    
    \item \emph{Network \ac{EM} theory of information: } There is the need of a full understanding of the fundamental performance limits as well as the development of practical algorithms for wireless networks composed of multiple users, base stations, scatterers, relays and \ac{RIS} under different geometrical configurations 
    \cite{FraMigMinSch:11,HuChiRusEdf:18,JunSaaJanKonChgo:19,JunSaaJanKoCho:20,BjoSan:20}. 

    \item \emph{Channel models:} One peculiarity of \acp{LIS} is that the communication channel may be no longer stationary along the surface and the \ac{EM} propagation may happen in the near-field condition where the wave front is spherical \cite{WuWanHaaAggAlwAi:15,PizMarSan:19}. 
    Ad hoc channel models should be developed and validated to account for such non-stationarity, including non-stationary polarization and the effect of multipath caused by near/far-field random scatterers.
    
    \item \emph{Channel estimation and localization:} The estimation of the \ac{CSI} is usually one of the most critical tasks in wireless communication. Moreover, when operating in the near-field, the channel is even more informative thus increasing the associated complexity in estimation
    \cite{DiRenzoJSAC:20,JunSaaKon:19}. 
    On the other hand, when moving at higher frequencies, obstacles may completely block the signal and multipath components become sparse so that communication is mainly enabled by \ac{LOS} conditions. As a consequence, the \ac{CSI} is expected to be highly correlated to the geometric configuration of antennas, i.e., the relative position and orientation, so that \ac{CSI} estimation and localization tasks become intimately linked and they can be tackled jointly. 
    

\end{itemize}

Holographic communication is an holistic way to manipulate the \ac{EM} field, generated or sensed by an antenna, with the maximum flexibility. 
Such flexibility can be enabled by new metamaterials designed for the realization of \acp{LIS} and \ac{RIS}. 
New wireless networks operating at millimeter wave and terahertz are expected to gain significant benefits in terms of increased capacity, reliability, and nodes densification.   
Perhaps, more than in the past, this process will require a tight synergy between the design at digital and \ac{EM} levels by leveraging the advances in \ac{EM} theory of information.

\ifCLASSOPTIONcaptionsoff
\fi
\bibliographystyle{IEEEtran}


\end{document}